\documentclass[preprint]{nature}

\usepackage{graphicx}%
\usepackage{multirow}%
\usepackage{amsmath,amssymb,amsfonts}%
\usepackage{amsthm}%
\usepackage{mathrsfs}%
\usepackage[title]{appendix}%
\usepackage{xcolor}%
\usepackage{textcomp}%
\usepackage{manyfoot}%
\usepackage{booktabs}%
\usepackage{algorithm}%
\usepackage{algorithmicx}%
\usepackage{algpseudocode}%
\usepackage{listings}%
\usepackage[mathlines]{lineno}
\usepackage{color}
\usepackage[10pt]{moresize}
\usepackage{amscd}
\usepackage{enumerate}
\usepackage{epsfig}
\usepackage{subfigure}
\usepackage{bm}
\usepackage[percent]{overpic}
\usepackage{graphics}
\usepackage{adjustbox}
\usepackage{braket}
\usepackage{caption}
\usepackage{epstopdf}
\usepackage{dcolumn}
\usepackage[colorlinks=true,citecolor=blue,urlcolor=black]{hyperref}

\begin{document}
\title{Experimental single-photon quantum key distribution surpassing the fundamental coherent-state rate limit}
\author
	{Yang Zhang$^{1,2,3,\ast}$, Xing Ding$^{1,2,3,\ast}$, Yang Li$^{1,2,3,\ast}$, Likang Zhang$^{1,2,3}$, Yong-Peng Guo$^{1,2,3}$, Gao-Qiang Wang$^{1,2,3}$, Zhen Ning$^{1,2,3}$, Mo-Chi Xu$^{1,2,3}$, Run-Ze Liu$^{1,2,3}$, Jun-Yi Zhao$^{1,2,3}$, Geng-Yan Zou$^{1,2,3}$, Hui Wang$^{1,2,3}$, Yuan Cao$^{1,2,3}$, Yu-Ming He$^{1,2,3}$, Cheng-Zhi Peng$^{1,2,3}$, Yong-Heng Huo$^{1,2,3}$, Sheng-Kai Liao$^{1,2,3}$, Chao-Yang Lu$^{1,2,3}$, Feihu Xu$^{1,2,3}$, Jian-Wei Pan$^{1,2,3}$}


\maketitle
\begin{affiliations}
	\item Hefei National Research Center for Physical Sciences at the Microscale and School of Physical Sciences, University of Science and Technology of China, Hefei 230026, China
	\item Shanghai Research Center for Quantum Science and CAS Center for Excellence in Quantum Information and Quantum Physics, University of Science and Technology of China, Shanghai 201315, China
	\item Hefei National Laboratory, University of Science and Technology of China, Hefei 230088, China
        \\
	$^\ast$These authors contributed equally.
\\
\end{affiliations}

\begin{abstract}
Single-photon sources are essential for quantum networks, enabling applications ranging from quantum key distribution (QKD) to the burgeoning quantum internet.
Despite the remarkable advancements, the current reliance of QKD on attenuated coherent (laser) light sources has imposed a fundamental limit on the secret key rate (SKR). This constraint is primarily attributable to the scarcity of single-photon components within coherent light, confined by an inherent upper bound of $1/{e}$. Here, we report high-rate QKD using a high-efficiency single-photon source, enabling an SKR transcending the fundamental rate limit of coherent light. We developed an on-demand, bright single-photon source with an efficiency of 0.71(2), exceeding the inherent bound of coherent light by approximately \textcolor{black}{2.87 dB}. Implementing narrow-bandwidth filtering and random polarization modulation, we conducted a field QKD trial over a 14.6(1.1)-dB-loss free-space urban channel, achieving an SKR of 1.08$\times10^{-3}$ bits per pulse. This surpasses the practical limit of coherent-light-based QKD by \textcolor{black}{2.53 dB}. Our findings conclusively demonstrate the superior performance of nanotechnology-based single-photon sources over coherent light for QKD applications, marking a pivotal stride towards the realization of a global quantum internet.
\end{abstract}
\maketitle

\paragraph{Introduction.}
The quantum internet\cite{kimble2008quantum,lu2021quantum} promises to connect distant nodes, enabling distributed quantum information processing. One privilege application of quantum internet is quantum key distribution (QKD), which promises fundamentally proven security for communication\cite{xu2020secure}. Since the first BB84 protocol\cite{Bennett1984}, significant progress has been made in extending QKD transmission distance and increasing secure key rates over optical fiber\cite{Peng2007RPL,rosenberg2007longPRL,lucamarini2018overcoming,Liu2023PRL,Li2023NP,Grunenfelder2023NP} and terrestrial free-space channels\cite{Manderbach2007PRL,Nauerth2013NP,Liao2017NP}. Satellite-to-ground QKD\cite{Liao2017Nature} and integrated space-ground quantum communication networks\cite{Chen2021Nature} have been extensively demonstrated.

To date, QKD experiments have primarily relied on attenuated laser sources\cite{Peng2007RPL,rosenberg2007longPRL,lucamarini2018overcoming,Liu2023PRL,Li2023NP,Grunenfelder2023NP,Manderbach2007PRL,Nauerth2013NP,Liao2017NP,Liao2017Nature,Chen2021Nature}.
Decoy-state QKD\cite{Lo2005PRL,Wang2005PRL}, in particular, has yielded impressive results, achieving secure key rates (SKR) exceeding 100 Mbps\cite{Li2023NP,Grunenfelder2023NP} and transmission distances up to 1000 km in fiber\cite{lucamarini2018overcoming,Liu2023PRL}. However, the Poissonian nature of weak coherent pulses (WCP) emitted from attenuated lasers imposes a fundamental limit on the SKR. For a phase-randomized WCP, the photon-number probability follows a Poisson distribution $P_{\left<n\right>}(k)= \left<n\right>^{k}e^{-\left<n\right>}/k!$, where $\left<n\right>$ is the mean photon number per pulse. The probability of single-photon components is inherently upper-bounded by $P_{\left<n\right>}(1)\leq 1/e$ when $\left<n\right>=1$. Consequently, for any WCP-based QKD system, the fundamental SKR achievable using single-photon components for key generation is limited by this bound\cite{Lo2005PRL,Wang2005PRL}.

In contrast, semiconductor quantum-dot single-photon source (SPS) can emit indistinguishable single photons on demand with a lower probability of producing multiple photons, offering a significant advantage for QKD\cite{Senellart2017NN,GarciadeArquer2021Science,Couteau2023NRP}. Ideal SPSs have an upper limit of 100\% for the single-photon probability, potentially enabling much higher SKRs than WCPs\cite{bozzio2022enhancing}. Several pioneering initiatives have explored SPS-based QKD experiments, demonstrating their promise in both fiber\cite{Takemoto2010APE,Takemoto2015SR,morrison2023single,zahidy2024quantum,yang2023high} and free-space channels\cite{Waks2002Nature,Heindel2012NJP,Rau2014NJP,Samaner2022AQT,Gao2023npj}. However, the lack of bright SPSs\cite{he2015single,wang2019towards,yu2023telecom} has hampered progress beyond the bound of WCPs, and the absolute advantage of SPSs in experiments remains elusive.

Here we demonstrate high-rate QKD which, for the first time, surpasses the fundamental rate limit of WCPs. Our on-demand, bright quantum-dot SPS\cite{ding2023high} exhibits an overall efficiency of 0.71(2), exceeding the inherent bound of coherent states of $1/{e}$ by \textcolor{black}{2.87 dB}. By employing narrow-linewidth filtering and active random polarization modulation, we show a 76.13 MHz SPS QKD light source with an overall efficiency of 0.292(8), $g^{(2)}(0)$ of 0.00698(64), and an encoding error of 2.54\%. In laboratory experiments, we report an SKR of 5.65$\times10^{-2}$ bit per pulse, representing an enhancement of 5.40 dB over WCP-based QKD under the same configuration. In a field trial, we implemented QKD over a 14.6(1.1)-dB loss free-space channel in an urban area at an SKR of 1.08$\times10^{-3}$ bits per pulse, which surpasses the key rate of WCP-based QKD by 2.53 dB. Our results demonstrate the superior advantage of SPS for QKD applications, paving the way for advanced single-emitter-based quantum communication and a future quantum internet.

\paragraph{The rate limit of QKD using WCP and SPS.}
We analyze the fundamental performance of WCP and SPS QKD in the asymptotic scenario, where the number of transmitted pulses is infinite. For WCP QKD, we consider the efficient BB84 protocol with infinite decoy states as the upper bound of the key rate\cite{Lo2005PRL,Wang2005PRL}.
In this scenario, the single-photon component can be accurately estimated, and the SKR is $R=\eta\left<n\right> e^{-\left<n\right>}$, where $\eta$ is the channel transmittance and $\left<n\right>$ is the average photon number per pulse. The maximum value of $R$ is achieved when $\left<n\right>=1$. The SKR of WCP-based QKD is thus given by
\begin{equation}
    R_{\text{WCP}}=\frac{1}{e}\eta
\end{equation}

For SPS-based QKD, we consider the BB84 protocol without decoy states, which aligns with our experiments. We follow the photon number distribution assumption $\{p_n\}=\{p_0,p_1,p_2\}$\cite{morrison2023single}, where the multi-photon contribution is dominated by the two-photon component. The SKR of SPS-based QKD is given by (see Supplemental Material),
\begin{equation}
R_{\text{SPS}} = 
\begin{cases}
-\dfrac{1}{2}g^{(2)}(0)\left<n\right>^2\left(\eta^2+1\right)+\left<n\right>\eta & \eta>\eta_{\text{th}} \\
\dfrac{\eta^2}{2g^{(2)}(0)\left(\eta^2+1\right)} & \eta \leq \eta_{\text{th}}, \\
\end{cases}
\end{equation}
where $\left<n\right>$ is the mean photon number, $g^{(2)}(0)$ is the second-order correlation function, and $\eta_{\text{th}}=(1-\sqrt{1-4\left(g^{(2)}(0)\left<n\right>\right)^2})/(2g^{(2)}(0)\left<n\right>)$.

\begin{figure*}[htbp]
\centering
\begin{overpic}[width=0.8\textwidth]{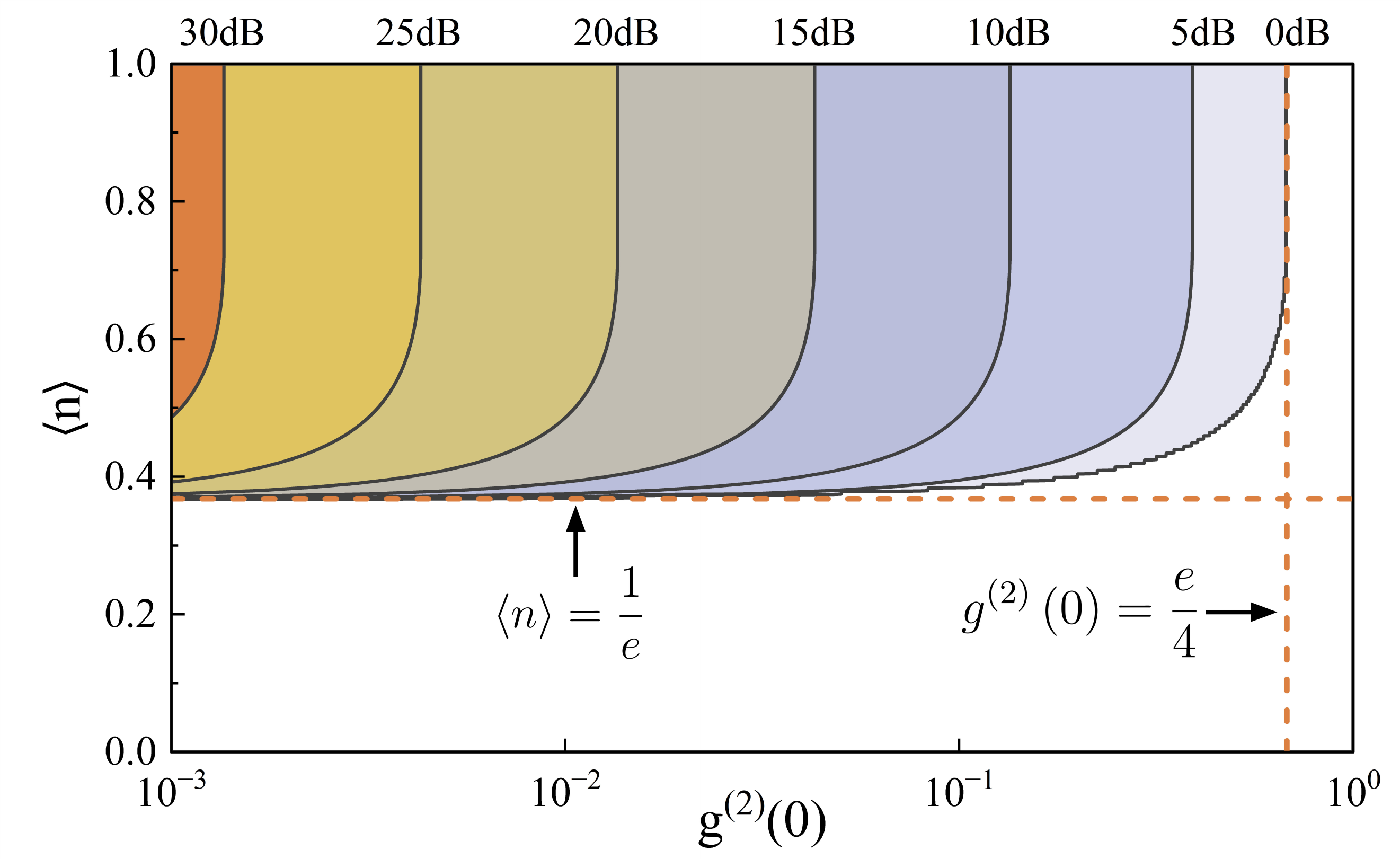}
\end{overpic}
\caption{
Configuration constraints enabling single-photon quantum key distribution (QKD) to surpass the rate limit of coherent states.
The black lines represent the lower bounds of the mean photon number $\left<n\right>$ and upper bounds of the second-order correlation function $g^{(2)}(0)$ for different channel losses of 0$\sim$30 dB.
The yellow dashed line indicates the fundamental bounds of $\langle n \rangle\geq 1/e~,~g^{(2)}(0)\leq e/4$ that SPS-based QKD must achieve to realize a key rate advantage.
}
\label{fig_bound}
\end{figure*}

The calculated configuration constraints required for single-photon QKD to surpass the rate limit of coherent states are illustrated in Fig.~\ref{fig_bound}.
Under varying channel losses, SPS QKD must reach large mean photon number and small second-order correlation function to outperform WCP QKD.
At zero channel loss, the fundamental bounds of the mean photon number $\left<n\right>$ and second-order correlation function $g^{(2)}(0)$ for SPS-based QKD are given by
\begin{equation}\label{equation_bound}
    \langle n \rangle\geq\frac{1}{e}~,~~ g^{(2)}(0)\leq\frac{e}{4}.
\end{equation}
As channel loss increases, the mean photon number required to achieve the key rate advantage bound gradually increases, while the second-order correlation function gradually decreases. This analysis provides the theoretical fundamental rate limit between WCP-based QKD and SPS-based QKD in an ideal system.

In a practical, finite-key scenario (see Supplemental Material), we employ the efficient three-state QKD protocol in our experiment\cite{xu2015experimental}, where $\ket{H}$ and $\ket{V}$ states on the Z basis and $\ket{R}$ state on the X basis are prepared.
For finite-key analysis, we utilize a tighter multiplicative Chernoff bound to constrain the estimated parameters, thereby achieving a higher secure key rate (SKR).
For a finite block size defined by the number of pulses sent $N_S$ or photons received $N_R$, the total security key length $\mathcal{L}$ is given by\cite{morrison2023single}:
\begin{align}
\mathcal{L} \geq \underline{N}^Z_{R,nmp}(1-H(\overline{\phi}^Z))-\lambda_{EC}-2\log_2\frac{1}{2\varepsilon_{PA}}-\log_2\frac{2}{\varepsilon_{cor}},
\end{align}
where $\underline{N}^Z_{R,nmp}$ represents the lower bound of the number of photons received on the basis of key generation (including both vacuum and emitted photons), $\overline{\phi}^Z$ denotes the upper bound of the phase error rate on the basis of key generation, $\lambda_{EC}$ is the information leaked during error correction, $\varepsilon_{PA}$ and $\varepsilon_{cor}$ are the secrecy and correctness parameters, respectively. The secure key rate per pulse is then calculated as $r=\mathcal{L}/N_S$.

\begin{figure}[!ht]
\centering
\centering
\begin{overpic}[width=1\textwidth]{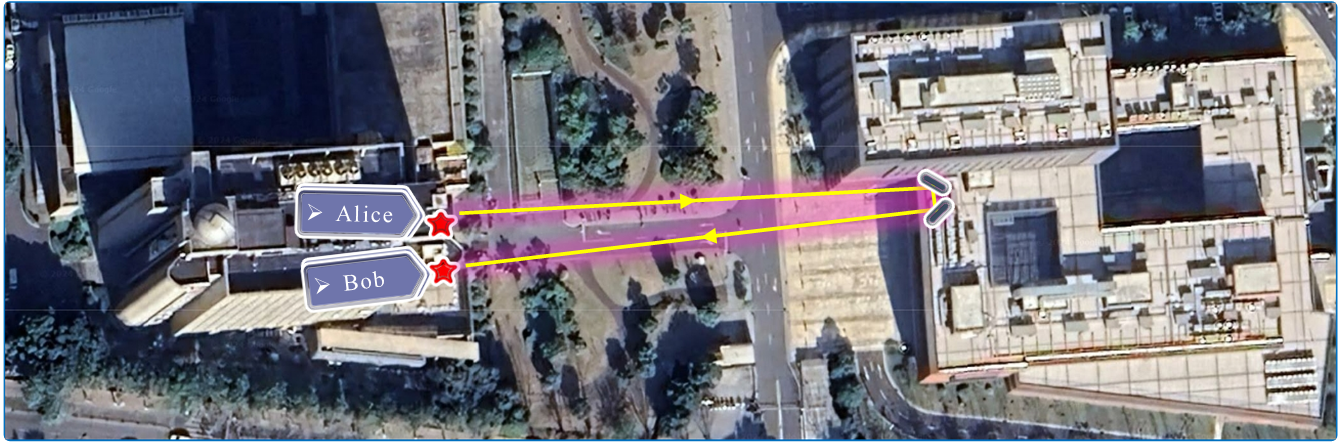}
\put(-4,31){\small\textbf{(a)}}
\end{overpic}
\hspace{2mm}
\begin{overpic}[width=1\textwidth]{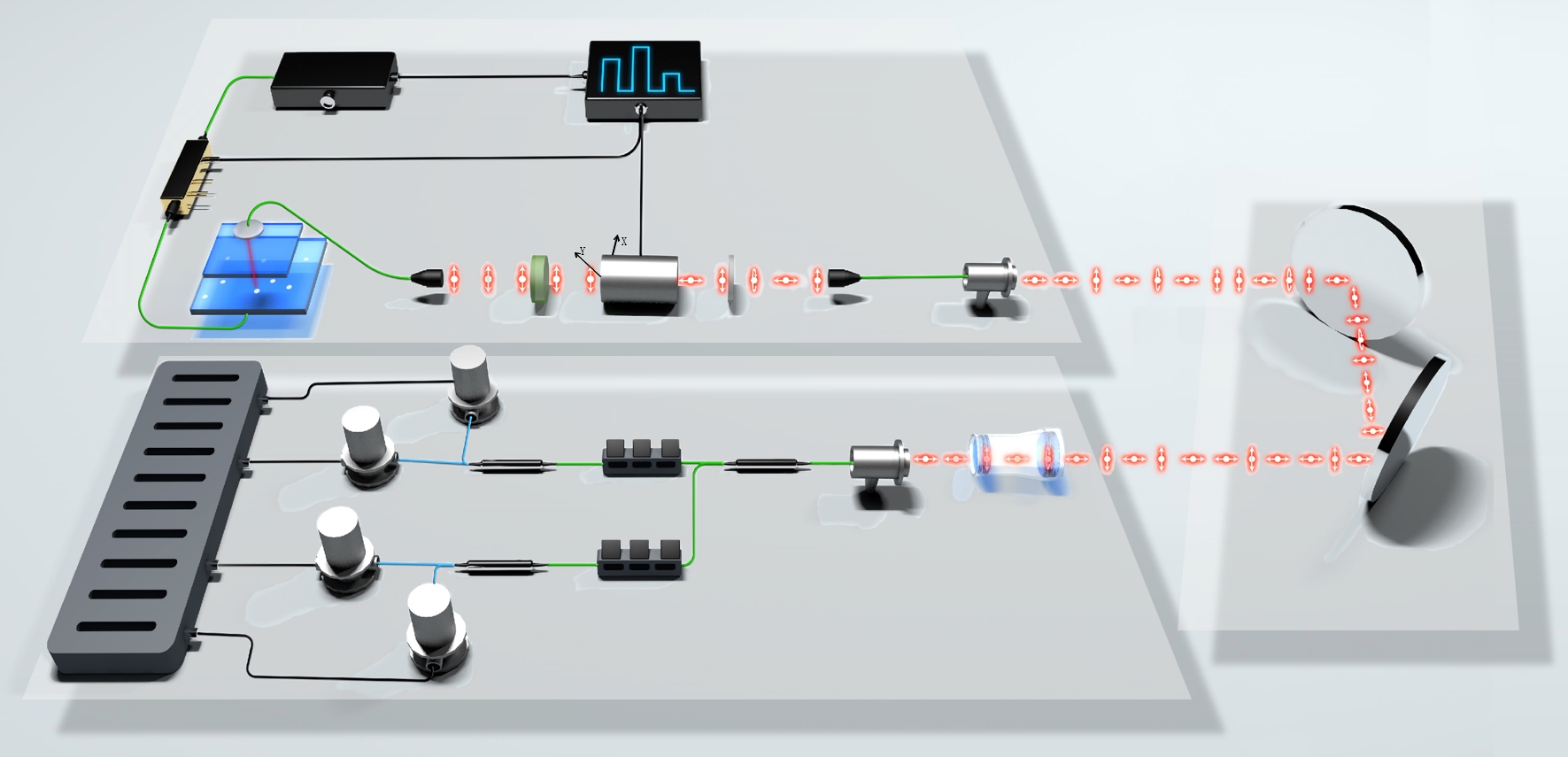}
\put(-4,46){\small\textbf{(b)}}
\put(38,46){\tiny\textbf{AWG}}
\put(17,46){\tiny\textbf{Excitation laser}}
\put(7,37){\tiny\textbf{PM}}
\put(16,27){\tiny\textbf{QD}}
\put(26,27){\tiny\textbf{Cpl}}
\put(33,27){\tiny\textbf{FP}}
\put(38.5,27){\tiny\textbf{EOM}}
\put(45,27){\tiny\textbf{ATT}}
\put(52,27){\tiny\textbf{Cpl}}
\put(62,27){\tiny\textbf{ET}}
\put(85,36){\tiny\textbf{RM}}
\put(64,15.5){\tiny\textbf{Col}}
\put(55,14.5){\tiny\textbf{RT}}
\put(47.8,16){\tiny\textbf{BS}}
\put(39.5,16){\tiny\textbf{PC}}
\put(39.5,10){\tiny\textbf{PC}}
\put(31,16){\tiny\textbf{PBS}}
\put(30,10){\tiny\textbf{PBS}}
\put(22,22){\tiny\textbf{SNSPD}}
\put(21.5,16){\tiny\textbf{SNSPD}}
\put(19,9){\tiny\textbf{SNSPD}}
\put(25,3.5){\tiny\textbf{SNSPD}}
\put(5,3.5){\tiny\textbf{TDC}}
\put(55,44.5){\small\textbf{Alice}}
\put(62,23.5){\small\textbf{Bob}}
\put(82,8.5){\small\textbf{Reflector}}
\end{overpic}
\caption{ 
\textbf{Overview of the single-photon QKD experiment.}
\textbf{a.} Birds-eye view of the QKD experiment, showcasing a free-space link between two co-located sites in Hefei.
The sites are connected via two 40 m optical fiber paths from the laboratory to each free-space launch point, enabling a 209 m air path.
\textbf{b.} The experimental setup.
The blue lines represent polarization-maintaining fibers, the green lines represent single-mode fibers, the black lines represent cables, and the red spherical symbols represent photons of different polarization states.
AWG, arbitrary wave generator; PM, phase modulator; QD, quantum dot; Cpl, coupler; FP, Fabry-Perot etalon; EOM, electro-optic modulator; ATT, attenuator; ER, emission telescope; RM, reflecting mirror; Col, collimator; RT, receiving relescope; BS, beam splitter; PC, polarization controller; PBS, polarization beam splitter; SNSPD, superconducting nanowire single-photon detector; TDC, time-to-digital converter.
}
\label{fig_setup}
\end{figure}

\paragraph{Experimental setup.}
As shown in Fig.~\ref{fig_setup}(a), the QKD experiment was conducted on the USTC campus in Hefei city (31$^\circ$49'59.4"N, 117$^\circ$16'10.8"E).
The QKD transmitter and receiver were co-located on the first floor of the Physicochemical building, connected by an optical fiber to their respective free-space launch sites for the subsequent 209-meter free-space transmission.

At the QKD transmitter (Fig.~\ref{fig_setup}b), high-efficiency single-photon source is realized by coupling a single InAs$\slash$GaAs quantum dot (QD) to a tunable plane-concave Fabry-Perot cavity, as detailed in the Supplementary Materials. A specific QD, exhibiting a 96\% quantum efficiency and an emission wavelength of 884.5 nm, is subsequently coupled to the open cavity. This cavity comprises a top concave mirror and a downward QD sample, both mounted on piezo nano-positioners. The entire QD-open microcavity system is cooled down to 4 K. A pulsed laser with a 69 GHz width and a repetition frequency of 76.13 MHz is used to resonantly excited the QD. Under $\pi$ pulse excitation, approximately 54 MHz single photons are collected into the single-mode fiber (SMF), resulting in an overall efficiency of 0.712(18) and $g^{(2)}(0)$ of 0.0205(6). This high-efficiency QD SPS, detailed in Reference\cite{ding2023high}, for the first time surpasses the fundamental rate limit of coherent states, a significant improvement over previous semiconductor QDs\cite{Takemoto2010APE,Takemoto2015SR,morrison2023single,zahidy2024quantum,yang2023high,Waks2002Nature,Heindel2012NJP,Rau2014NJP,Samaner2022AQT,Gao2023npj}.

To prevent potential phase coherence between consecutively emitted photons\cite{bozzio2022enhancing,karli2024controlling}, we implement phase randomization by introducing active phase modulations on the pump laser pulses. A Fabry-Perot etalon with a 5.4 GHz linewidth is employed to further filter out the background of the excitation laser from the emitted single photons. A free-space Pockels cell with a transmittance of 97\% is used to implement random polarization state modulation based on a predetermined random number sequence. We incorporate an adjustable attenuator into the optical path and employ the pre-attenuation technique\cite{morrison2023single} to enhance the SKR in scenarios involving large channel losses. Utilizing the connecting optical fiber and emission telescope, the prepared quantum photons are ultimately transmitted into the free-space channel. After implementing narrow-lindwidth filtering and random polarization modulation, we obtain a SPS QKD light source with an overall efficiency of 0.292(8), $g^{(2)}(0)$ of 0.00698(64), and a polarization encoding error of 2.54\% (see Supplemental Material).

At the QKD receiver (Fig.~\ref{fig_setup}d), photons are collected into the connecting optical fiber and then fed into the polarization decoding module, which comprises a 9:1 beam splitter (BS), two polarization controllers (PCs), and two polarization beam splitters (PBSs).
The polarization-decoded quantum photons are detected by four superconducting-nanowire single-photon detectors (SNSPDs) with detection efficiencies of approximately 0.712 and total dark counts of around 43 cps. A time-to-digital converter (TDC) with 42 ps resolution measures the arrival time of the detected photons.
Utilizing a quantum photon-based synchronization method\cite{Wang2021OE}, precise time synchronization between the two QKD terminals is realized, simplifying additional hardware and reducing complexity. A temporal filtering gate width of 3.42 ns is further employed to mitigate the effects of detector dark counts and urban environment background noise.

Detailed system and security parameters of the field QKD experiment are presented in Table~\ref{tab1_key}. In addition to the field experiment, we have also conducted several desktop experiments with different channel losses in the laboratory. Details of the QKD system parameters for the laboratory experiments are provided in the Supplemental Material. Notably, for different experiments, a genetic algorithm was used to numerically optimize the quantum state preparation ratio of different basis and the pre-attenuation applied at the QKD transmitter.

\begin{figure*}[htbp]
\centering
\hspace{2mm}
\begin{overpic}[width=0.9\textwidth]{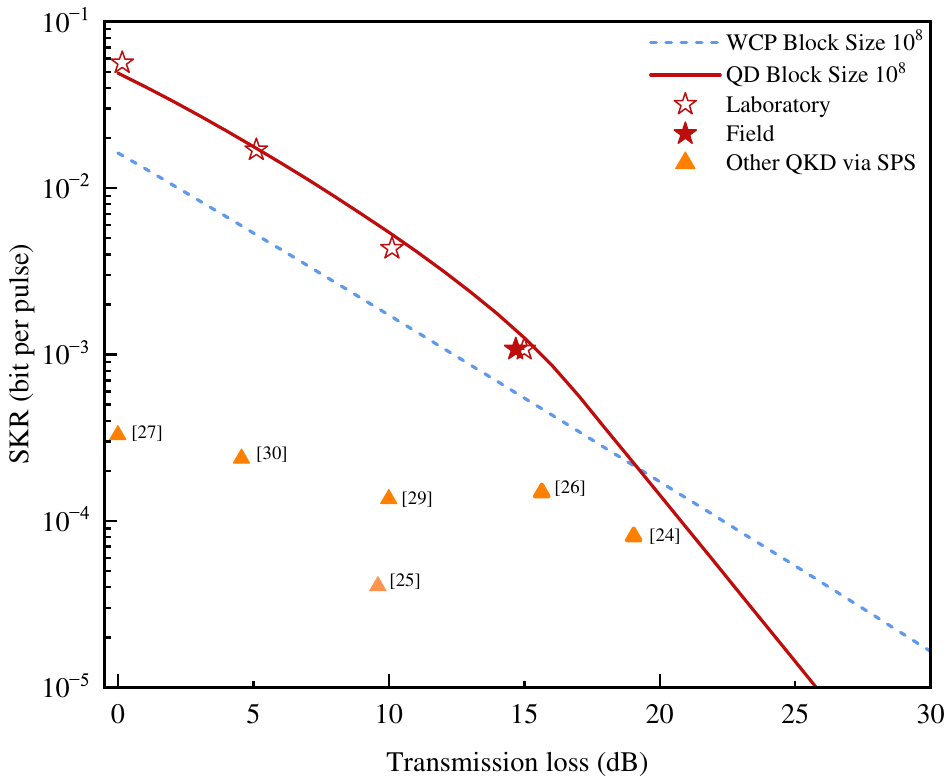}
\end{overpic}
\caption{
Results of single-photon QKD experiments and comparison to other experiments.
The solid five-pointed star represents the field experiment result, exceeding the key rate of WCP-based QKD by approximately \textcolor{black}{2.53 dB}.
The hollow five-pointed stars depict results from several laboratory experiments with varying channel losses, demonstrating a key rate improvement of approximately 2.95$\sim$5.40 dB over WCP-based QKD.
The solid triangles represent experimental results from other SPS-based QKD.
WCP, weak coherent pulse; SPS, single-photon source.
}
\label{SKR}
\end{figure*}

\paragraph{Results.}
We conducted a series of laboratory and field experiments with varying channel losses, with the security parameters shown in Table~\ref{tab1_key} and a finite block size of $10^8$ chosen by receiver for key distillation.
The results of our single-photon QKD experiments and comparisons with other experiments are depicted in Fig.~\ref{SKR}.
In the field experiment, the measured SKR is 82.0 kbps under an average channel loss of 14.6(1.1) dB.
Considering the repetition frequency of 76.13 MHz, the corresponding SKR is approximately 1.08$\times10^{-3}$ bits per pulse.
In laboratory experiments, the measured SKRs are 4.30 Mcps, 1.29 Mcps, 330 kcps, and 82.5 kcps for channel losses of 0.17(7) dB, 5.11(5) dB, 10.15(12) dB, and 15.16(10) dB, respectively.
The corresponding SKRs are approximately $5.65\times10^{-2}, 1.69\times10^{-2}, 4.34\times10^{-3}$, and $ 1.08\times10^{-3}$ bits per pulse, respectively.
It should be noted that the measured mean photon number and second-order correlation function slightly deviate from the results obtained in the field experiment.

Using the same system and security parameters as the field experiment (Table \ref{tab1_key}), we also conducted numerical simulations for both SPS-based QKD and WCP-based QKD. For WCP-based QKD, the average photon number, the ratio of each intensity state (signal state, decoy state, and vacuum state), and the quantum state preparation ratio of different basis, were all optimized numerically.
According to our simulation results, the key rate of SPS-based QKD exceeds that of WCP-based QKD under channel loss below 19 dB, with a maximum key rate improvement of approximately \textcolor{black}{5.40 dB} under near-zero loss. 
Our experiment results are consistent with the simulation results.
Compared to the theoretically expected key rate of WCP-based QKD, we achieved key rate enhancements of \textcolor{black}{2.95$\sim$5.40 dB and 2.53 dB} in the laboratory and field experiments, respectively. 
Furthermore, due to its inherent advantage in photon number distribution, SPS-based QKD is more robust to statistical fluctuations than WCP-based QKD (see Supplemental Material). 
Compared to other SPS-based QKD experiments\cite{Takemoto2015SR,morrison2023single,zahidy2024quantum,yang2023high,Waks2002Nature,Rau2014NJP,Samaner2022AQT}, we achieved about an order of magnitude improvement in the key rate under the same channel loss (see Supplemental Material).

\begin{figure*}[htbp]
\centering
\begin{overpic}[width=0.8\textwidth]{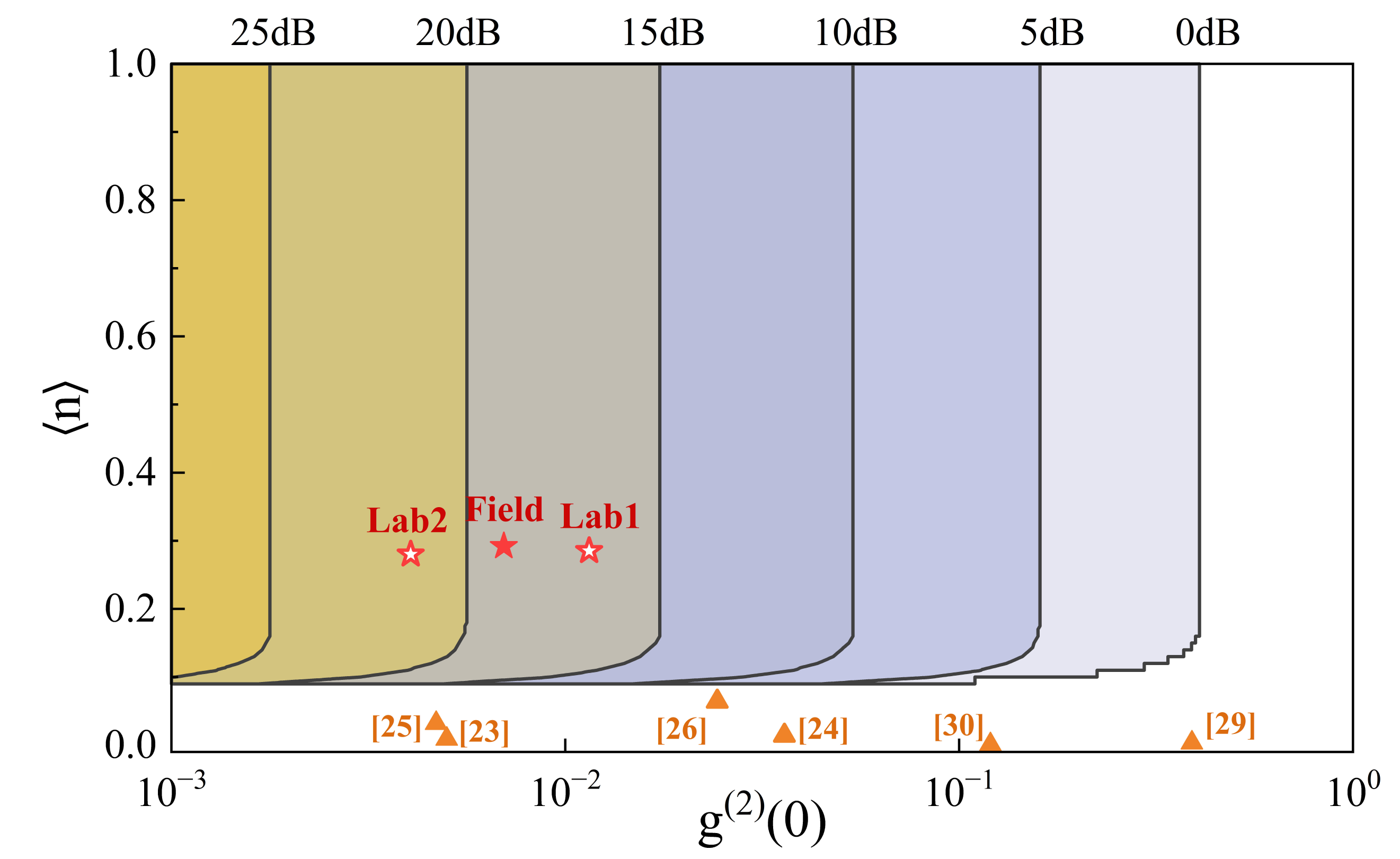}
\put(3,59){\small\textbf{(a)}}
\end{overpic}
\hspace{2mm}
\begin{overpic}[width=0.8\textwidth]{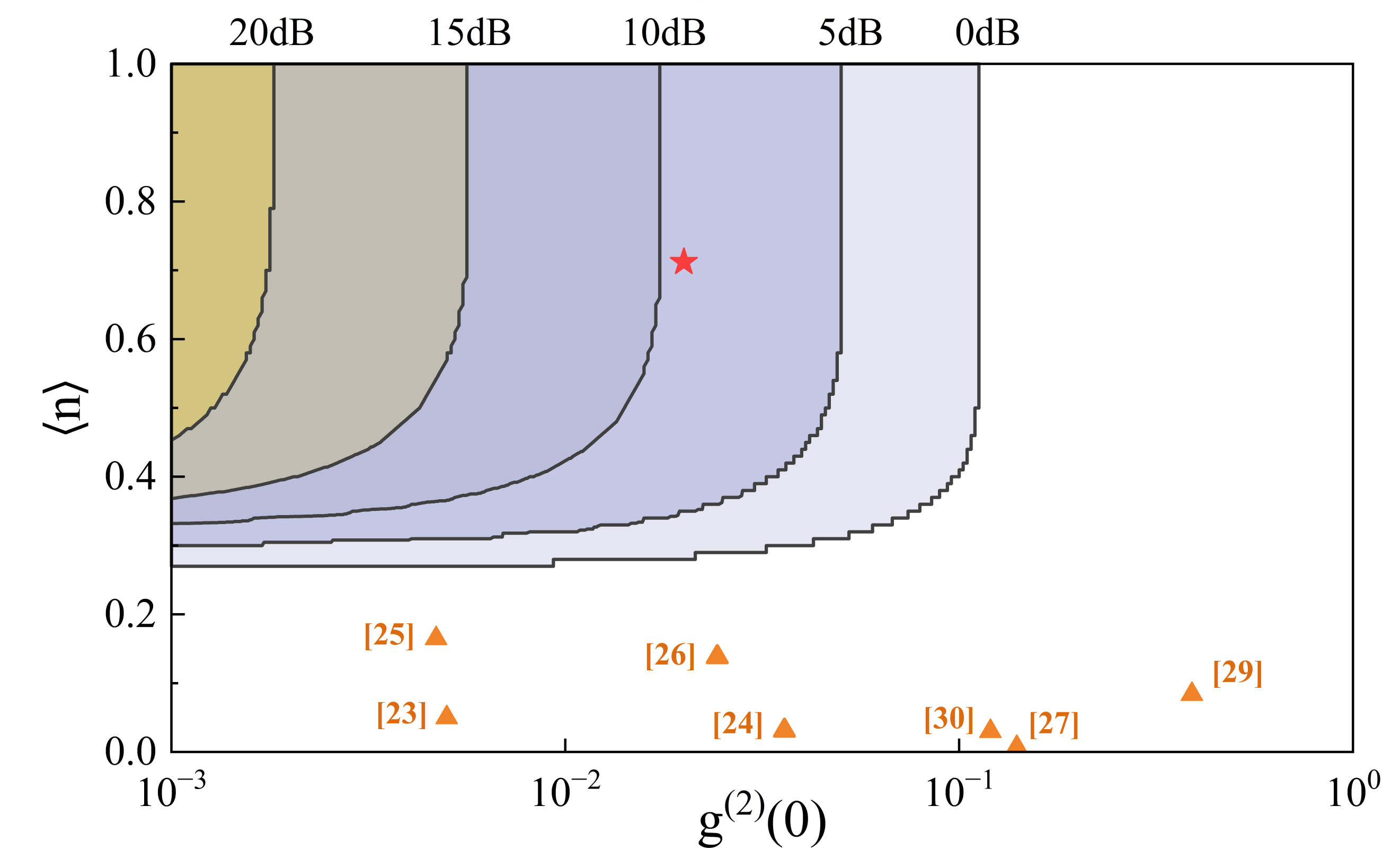}
\put(3,59){\small\textbf{(b)}}
\end{overpic}
\caption{
\textbf{a.} Configuration constraints for single-photon QKD to transcend the rate limit of coherent states using the same protocol, system and security parameters as the field experiment with a finite block size of $10^8$.
\textbf{b.} Configuration constraints for single-photon QKD to transcend the rate limit of coherent states under the asymptotic key regime. The solid five-pointed star represents the results of the field experiment, while hollow five-pointed stars represent several laboratory experiments. The solid triangle represents results from other SPS-based QKD.
}
\label{beat}
\end{figure*}

Following the same protocol, system, and security parameters as the field experiment (Table \ref{tab1_key}), we re-compared the performance of SPS and WCP, with the results shown in Fig. \ref{beat}(a).
The black solid line represents the fundamental bounds required for SPS QKD to achieve a key rate advantage, with a mean photon number $\left<n\right>$ of approximately 0.078 and a second-order correlation function $g^{(2)}(0)$ of approximately 0.41.
Note that these fundamental bounds are specific to the system parameters, such as the misalignment probability of 2.54\% and the finite block size of $10^8$, and are not absolute.
Considering ideal quantum state modulation and the asymptotic key regime, we can more fairly compare the performance of SPS with WCP for QKD, with the results shown in Fig. \ref{beat}(b).
The fundamental bounds for SPS QKD to achieve a key rate advantage shift to a mean photon number $\left<n\right>$ of approximately 0.268 and a second-order correlation function $g^{(2)}(0)$ of approximately 0.11.
In both cases, we experimentally demonstrate the key rate advantage of SPS-based QKD over WCP-based QKD, which has never been demonstrated in previous works.

\paragraph{Discussion.}
The 76.13 MHz repetition frequency of SPS-based QKD can be further increased to several gigahertz through techniques such as frequency doubling, comparable to state-of-the-art values of WCP-based QKD\cite{Li2023NP,Grunenfelder2023NP}. Another challenge for SPS-based QKD is the maximum tolerable channel loss, which is currently lower than the state-of-the-art attainable channel loss of WCP-based QKD\cite{boaron2018secure}. This limitation stems from the presence of multi-photon components in SPSs, which can negative impact under high channel loss conditions. This limitation can be addressed through two primary strategies: First, continuous efforts to improve SPS performance are needed to reduce multi-photon components\cite{schweickert2018demand}, where O/C band QD SPSs are now experiencing rapid developments\cite{yu2023telecom} so that long-distance operation is promising. Second, integrating decoy-state theory can establish a precise boundary for single-photon components and enhance the resilience against high losses\cite{Lo2005PRL,Wang2005PRL}. Additionally, although the present SPS operates at visible wavelengths, frequency conversion techniques can be explored to translate to the telecommunication band\cite{morrison2023single,2023Quantum,zahidy2024quantum}. Furthermore, it is worth investigating the potential applications of SPSs beyond BB84\cite{pirandola2020advances}, such as entanglement-based QKD\cite{liu2019solid,basso2021quantum}, measurement-device-independent QKD\cite{Lo2012} and twin-field QKD\cite{lucamarini2018overcoming}. Further efforts can be devoted to developing sophisticated quantum network infrastructures\cite{kimble2008quantum,lu2021quantum,thomas2022efficient} leveraging quantum teleportation\cite{bouwmeester1997experimental} and quantum repeaters\cite{briegel1998quantum}.

In summary, we have theoretically compared the performance of WCP and SPS for QKD, and experimentally demonstrated single-photon QKD that surpasses the fundamental rate limit of coherent states. Our results conclusively demonstrate the advantage of nanotechnology-based single-photon sources in QKD applications and serve as a strong foundation for future applications in practical quantum networks.

\paragraph{Data availability.}
The datasets generated in the current study are available from the corresponding author upon reasonable request.
	
\paragraph{Acknowledgments.}
We would like to thank Jin Liu, Zhen Ning, Ji-Gang Ren, Chao-Ze Wang, Xin-Zhe Wang, Juan Yin for helpful discussions and assistance. This work was supported by National Key Research and Development Program of China (2020YFA0309700, 2022YFF0610100), National Natural Science Foundation of China (62031024, 12174374, 12274398, 12374475), Innovation Program for Quantum Science and Technology (2021ZD0300104, 2021ZD0300300), Shanghai Municipal Science and Technology Major Project (2019SHZDZX01), Shanghai Science and Technology Development Funds (22JC1402900), Jinan Innovation Zone, Key Research and Development Plan of Shandong Province (2021ZDPT01) and the New Cornerstone Science Foundation through the Xplorer Prize.

\paragraph{Competing interests}
The authors declare no competing interests.

\paragraph{References}
\bibliographystyle{naturemag_noURL}
\bibliography{reference.bib}
\newpage
\clearpage

\newpage
\clearpage

\newpage
\clearpage

\newpage
\clearpage

\newpage
\clearpage

\begin{table*}
\centering
\caption{
System and security parameters for the field QKD experiment.
}\label{tab1_key}
\begin{tabular*}{\textwidth}{@{\extracolsep\fill}ccc}
\hline
Description & Parameter & Value \\
\hline
Clock rate & $R$ & $76.13$ MHz\\
Single photon pulse width & $\Delta T$ & $75.8$ ps\\
Single photon line width & $\Delta \nu$ & $2.73$ GHz\\
Single photon collection efficiency & $\eta_{QD}$ & $0.71$\\
Transmitter efficiency & $\eta_{T}$ & $0.410$\\
Mean photon number & $\langle n \rangle$ & $0.292$\\
Second-order correlation & $g^{(2)}(0)$ & $0.00698$\\
\hline
Channel loss & $l$ & $-14.6$ dB\\
Fiber optics efficiency & $\eta_{FO}$ & $0.6$\\
Detection efficiency & $\eta_{D}$ & $0.712$\\
Detector dark count & $DC$ & $43$ cps\\
Temporal filtering gate width & $t_{GW}$ & $3.42$ ns\\
Dark count probability & $p_{DC}$ & $1.47\times10^{-7}$\\
Misalignment probability & $p_{mis}$ & $2.54\%$\\
\hline
Parameter estimation failure probability & $\varepsilon_{PE}$ & $11\times10^{-10}/12$\\
Privacy amplification failure probability & $\varepsilon_{PA}$ & $10^{-10}/24$\\
Error correction failure probability & $\varepsilon_{EC}$ & $10^{-10}/24$\\
Correctness failure probability & $\varepsilon_{cor}$ & $10^{-15}$\\ 
Error correction efficiency factor & $f_{EC}$ & $1.16$\\
\hline
\end{tabular*}
\end{table*}
\newpage
\clearpage

\end{document}